\documentclass[prl,10pt,floatfix,amsmath,twocolumn,showpacs,superscriptaddress]{revtex4-2}
\usepackage{latexsym}
\usepackage{amsmath}
\usepackage{amssymb}
\usepackage[ansinew]{inputenc}
\usepackage{graphicx}
\usepackage{txfonts}
\usepackage{booktabs}
\usepackage[colorlinks=true,linkcolor=blue,
pagecolor=blue,filecolor=blue,
menucolor=blue,urlcolor=blue,
citecolor=blue,anchorcolor=blue]{hyperref}

\begin{document}

\title{Prediction of topological superconductivity in 1$T$-TiTe$_2$ under pressure}

\author{Wenjun Liu}
\affiliation{Wuhan National High Magnetic Field Center and School of Physics,
Huazhong University of Science and Technology, Wuhan, 430074, China}

\author{Aiyun Luo}
\affiliation{Wuhan National High Magnetic Field Center and School of Physics,
Huazhong University of Science and Technology, Wuhan, 430074, China}

\author{Guyue Zhong}
\affiliation{Wuhan National High Magnetic Field Center and School of Physics,
Huazhong University of Science and Technology, Wuhan, 430074, China}

\author{Jinyu Zou}
\affiliation{Wuhan National High Magnetic Field Center and School of Physics,
Huazhong University of Science and Technology, Wuhan, 430074, China}

\author{Gang Xu}
\email{gangxu@hust.edu.cn}
\affiliation{Wuhan National High Magnetic Field Center and School of Physics,
Huazhong University of Science and Technology, Wuhan, 430074, China}

\begin{abstract}
Topological superconductivity has attracted intensive interest for its ability of hosting Majorana zero mode and implementing in topological quantum computations.
Based on the first-principles calculations and the analysis of the effective BdG Hamiltonian,
we demonstrate that 1$T$-TiTe$_2$ is a topological metal hosting Dirac cone type of surface states near the Fermi level,
and it exhibits a normal-topological-normal superconducting phase transition as a function of the chemical potential.
These results point out a new promising topological superconductor without random dopant, in which the influence of the impurity may be greatly reduced.
Furthermore, our calculations also suggest that the transition metal intercalated Ti(Se$_{1-y}$Te$_y$)$_2$ is also a highly possible route to realize topological superconductivity and Majorana zero modes.
\end{abstract}

\maketitle

\emph{Introduction} ---
As one of the most intriguing systems to host the Majorana zero modes (MZMs), topological superconductors (TSCs) play important roles in both the condensed matter physics and the topological quantum computations~\cite{Kitaev_2001, KITAEV20032, PhysRevB.61.10267}, which have attracted growing interest in the past decade~\cite{Sarma2015, Sato_2017, Lutchyn2018}.
Since natural TSC is very rare~\cite{Mackenzie2017}, a variety of architectures have been proposed in one-, two- and three-dimension for the realization of the TSC and MZMs
~\cite{sarma2006, Nayak2008, PhysRevLett.100.096407, PhysRevLett.103.237001, PhysRevA.82.052322, PhysRevB.82.094522, PhysRevLett.105.077001, PhysRevLett.105.177002, PhysRevB.81.125318, PhysRevLett.105.227003, Nat.Phys.7.412, PhysRevLett.105.227003, PhysRevB.83.054513,weng2011, xu2014, PhysRevLett.117.047001, zou2019, Pan2019, chen2019}.
In particular, Gang Xu \emph{et al.} proposed that TSC can be realized
in the vortex of superconducting topological metallic materials such as the iron-based superconductor FeSe$_{0.5}$Te$_{0.5}$.
Recently, the MZMs and the superconducting gap on the surface states are observed in Fe(Se,Te)~\cite{Yin2015, Machida2019, nphys1745} and (Li$_{0.84}$Fe$_{0.16}$)OHFeSe~\cite{PhysRevX.8.041056}.
Such exiting progresses promote the superconducting topological metallic materials to the forefront for the exploring of the MZMs and the topological qubits~\cite{Nat.Phys.7.412, PhysRevLett.105.077001}.

However, the MZMs on the surface of the iron-based superconductors seems delicate~\cite{zhang2018, ChenMingyang2018}.
Different annealing treatment can even eliminate the MZMs, and lead to the discrete Caroli-de Gennes-Matricon states~\cite{ChenMingyang2018}.
Chen \emph{et al.} have attributed such observation to the multi-band character of the iron-based superconductors.
But the conclusive explanations for these contradictory experimental results are still lacking and under debate~\cite{wang2018evidence, wang2020evidence, fan2021observation}.
The other vital obstacle is the inhomogeneous electronic properties caused by the multiple random dopants, which has been proven by the spatial inhomogeneity of the tunneling
spectra~\cite{ChenMingyang2018}. Therefore, it is highly desirable to search for new superconductors that host the topological electronic structures at the Fermi level but have no random doping.
So that one can study the survival conditions of the MZMs in the vortex more clear and realize the more robust MZMs.

In this paper, by means of the first-principles calculations, we demonstrate that the layered transition-metal dichalcogenide 1$T$-TiTe$_2$ is a superconducting topological metal.
Its band structures manifest a $p-d$ band inversion and a topologically nontrivial gap near the Fermi level.
Dirac cone type of the topological surface states could be stabilized on the (001) surface.
When 1$T$-TiTe$_2$ enters the superconducting phase under about 5$-$12~GPa~\cite{PhysRevB.97.060503},
the calculated energy spectrum of the vortex BdG Hamiltonian exhibits a normal-topological-normal superconducting phase transition.
These results confirm that 1$T$-TiTe$_2$ is a promising TSC candidate.
Because there is no random dopant in 1$T$-TiTe$_2$, the TSC phase in it could be more homogeneous.
Further studies suggest that the transition metal intercalated Ti(Se$_{1-y}$Te$_y$)$_2$ is also a highly possible route to realize TSC and MZMs.

\begin{figure}[!t]
\includegraphics[width=0.5\textwidth]{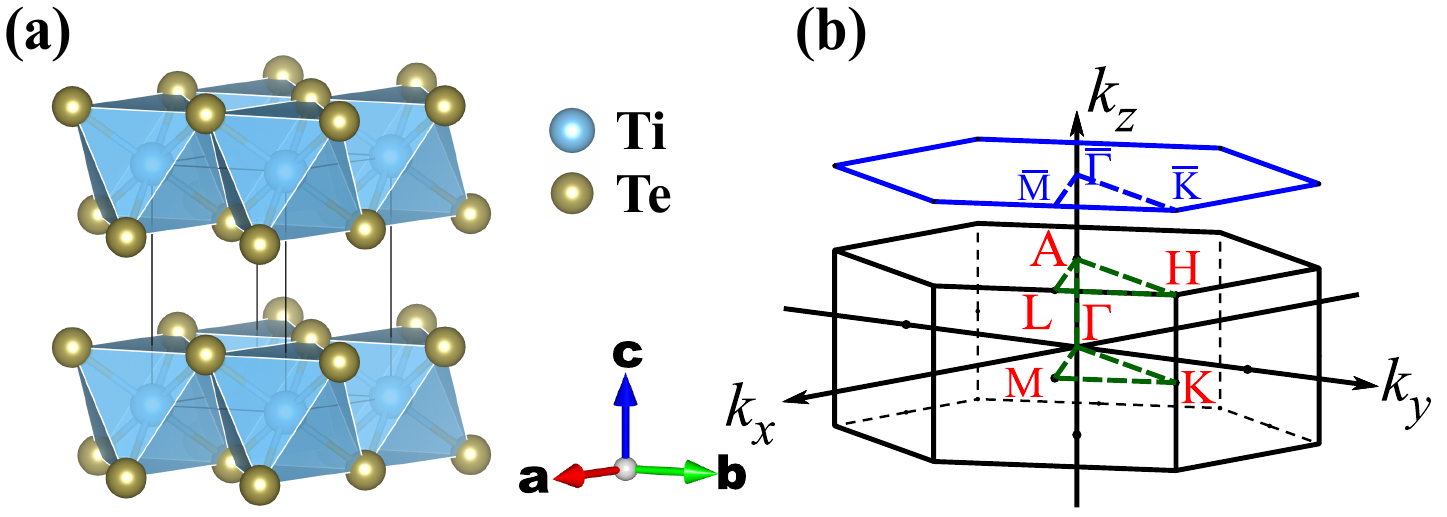}
\caption{
    (a) The crystal structure of 1$T$-TiTe$_2$ in space group $P\bar{3}m1$.
    (b) High symmetry $k$-point in the bulk Brillouin zone and their projection on the (001) surface.
}
\label{fig:fig1}
\end{figure}

\emph{Crystal structure and methodology} ---
As shown in Fig.~\ref{fig:fig1}(a), 1$T$-TiTe$_2$ adopts a triangular lattice structure with space group $P\bar{3}m1$ (No.~164)~\cite{ARNAUD1981230},
in which each layer of Ti atom is surrounded by two layers of Te atoms, and form a stable TiTe$_2$-octahedral sandwich.
The TiTe$_2$-octahedral sandwiches are A-A stacked along the $c$-axis, and finally construct a layered Van-der Waals material.
It undergoes a charge density wave transition at room temperature~\cite{NatComm.8.516, TiTe2CDW}.
When an external pressure of about 5$-$12~GPa is applied, the charge density wave transition is suppressed,
and a superconducting phase with highest $T_c$ = 5.3~K is induced~\cite{PhysRevB.97.060503}.
The optimized crystal parameters $a = 3.85$~{\AA}, $c = 6.64$~{\AA} and $z =$ 0.245, which agree with the experimental results well, are used in our calculations.
Our first-principles calculations are performed by the Vienna \emph{ab initio} simulation package~\cite{PhysRevB.54.11169, kresse1996efficiency}
based on the density functional theory (DFT) and projected augmented wave method ~\cite{PhysRevB.50.17953}.
The energy cutoff is set as 400 eV, and $11\times11\times9$ k-meshes are adopted. Perdew-Burke-Ernzerhof type of the exchange-correlation potential~\cite{PhysRevLett.77.3865},
and Heyd-Scuseria-Ernzerhof (HSE06) hybrid functional~\cite{heyd2003hybrid} with Hartee-Fock exchange factor 0.2 are used in all calculations to obtain the accurate electronic structures.
Spin-orbit coupling (SOC) interaction is considered consistently.

\begin{figure}
\includegraphics[width=0.49\textwidth]{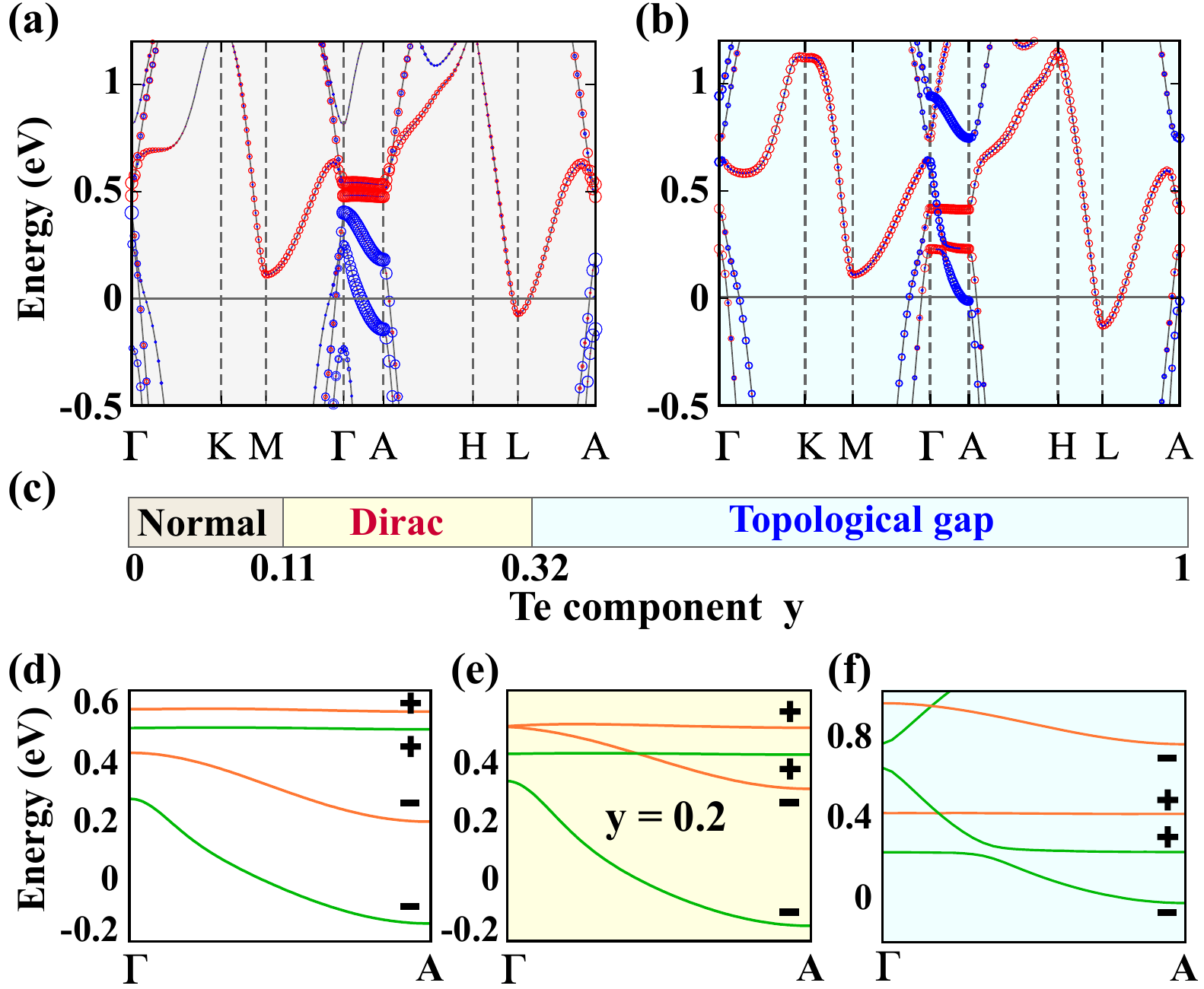}
\caption{
Band structures of 1$T$-TiSe$_2$  (a) and 1$T$-TiTe$_2$ (b) from the HSE06 calculations.
The size of red and blue circles represent the weight projections for the $d$ orbitals of Ti atoms and the $p$ orbitals of chalcogen atoms, respectively.
(c) The topological phase diagram of Ti(Se$_{1-y}$Te$_{y}$)$_2$.
(d) -- (f) are the band structures along $\Gamma - A$ path of TiSe$_2$, Ti(Se$_{0.8}$Te$_{0.2}$)$_2$, and TiTe$_2$, respectively,
in which we use orange and green bands to denote the $j^{~z}_{eff} = \pm3/2$ states and $j^{~z}_{eff} = \pm1/2$ states.
The even/odd (+/-) parity for each band at the $A$ point is also marked in (d) -- (f).
}
\label{fig:fig2}
\end{figure}

\emph{The first-principles calculations} ---
In Figs.~\ref{fig:fig2}(a) and~\ref{fig:fig2}(b), we calculate and compare the band structures of 1$T$-TiSe$_2$ and 1$T$-TiTe$_2$ respectively.
For both of them, two valence bands are not fully occupied, leading to two hole Fermi pockets around the $\Gamma-A$ path,
and one conduction band is partially occupied, leading to one electron Fermi pocket around the $L$ point.
Such semi-metallic characteristics agree well with many ARPES observations ~\cite{PhysRevB.63.125104, PhysRevB.72.085418, PhysRevB.99.125112},
which confirm the validity of our calculation method.
The orbital projections illustrate that the valance bands are mostly contributed by the $p$ orbitals ($p_{x} \& p_{y}$) of the chalcogen atoms,
while the conduction bands are mostly from the $d$ orbitals ($d_{z^2}$ around $L$ point) of Ti atoms.
These results are consistent with previous analysis that charge transfer happens from titanium to chalcogen,
and verify that the Ti ion has a chemical valence close to $+4$~\cite{PhysRevLett.110.077202}.
Moreover, we notice that two conduction bands with even parity contributed by the $d$ orbitals ($d_{xz} \& d_{yz}$) are just
78 meV higher than the valence band top (odd parity) at $\Gamma$ in the 1$T$-TiSe$_2$, as shown in Fig.~\ref{fig:fig2}(a).
If the energy of the $d$ bands can be modulated a little lower, $p-d$ band inversion could occurs along the $\Gamma-A$ path.
This is exactly what happened in 1$T$-TiTe$_2$.
As shown in Fig.~\ref{fig:fig2}(b), the $p-d$ band overlap is enhanced drastically when the Se atoms are substitute by the Te atoms.
As a result, one upper $p$ band is pushed above the $d$ bands along the $\Gamma-A$ path,
and one lower $p$ band intersects with the $d_{xz} \& d_{yz}$ bands,
forming the $p-d$ band inversion along the $\Gamma-A$ path.
Therefore, 1$T$-TiTe$_2$ becomes a topological metal~\cite{zhang2018preesure},
which is very similar to the band structures of the Fe(Se,Te)~\cite{PhysRevLett.117.047001, PhysRevB.92.115119}.

These results suggest that the topological metal phase can be engineered by doping Te to 1$T$-TiSe$_2$~\cite{PhysRevLett.110.077202}.
We thus perform the virtual crystal approximation calculations on the band structures of Ti(Se$_{1-y}$Te$_y$)$_2$.
The calculated results give rise to three phases depending on the Te component $y$, as shown in Fig.~\ref{fig:fig2}(c).
When $y$ is less than 0.11, the $d$ bands with even parity are always higher than the $p$ bands with odd parity,
and the system falls into the normal metal phase as represented in Fig.~\ref{fig:fig2}(a) and Fig.~\ref{fig:fig2}(d).
As Te component $y$ increasing, the chemical bonding between the $p$ and $d$ orbitals is weakened
because the electronegativity of Te is smaller than that of Se.
Consequently, the $d$ bands are pushed lower and lower, and the system enters the Dirac semimetal phase at 0.11$< y <$ 0.32,
as represented by the band structures of Ti(Se$_{0.8}$Te$_{0.2}$)$_2$ in Fig.~\ref{fig:fig2}(e).
When $y$ is large than 0.32, the system becomes a topological metal with a nontrivial band gap near the Fermi level as shown in Fig.~\ref{fig:fig2}(b) and Fig.~\ref{fig:fig2}(f).

To understand the whole phase diagram in Fig.~\ref{fig:fig2}(c), we should take into account the SOC interaction and analyse the $C_{3z}$ irreducible representation of each band along the $\Gamma-A$ path.
As shown in Figs.~\ref{fig:fig2}(d)-(f), our calculations demonstrate that the ($p_{x} \& p_{y}$) orbitals and ($d_{xz} \& d_{yz}$) orbitals always split into the same $j^{~z}_{eff}$ order by SOC interaction,
i.e., $j^{~z}_{eff}=\pm3/2$ states [orange bands in Figs.~\ref{fig:fig2}(d)-(f)] are higher than the $j^{~z}_{eff}=\pm1/2$ states [green bands in Figs.~\ref{fig:fig2}(d)-(f)] for both $p$ and $d$ orbitals.
This seems contradictory to the conventional knowledge that $d$ orbitals have the negative SOC interaction with respect to that of the $p$ orbitals~\cite{Shindo1965, PhysRevLett.106.236806, Sheng2017}.
But it is a reasonable result due to the strong $p-d$ hybridization that has been confirmed in many other materials~\cite{PhysRevB.12.3330, PhysRevB.30.5904, PhysRevB.90.245308}.
Once the ordering is aligned as in Figs.~\ref{fig:fig2}(d)-(f), the $d_{\pm 1/2}$ states will invert with the $p_{\pm 3/2}$ states firstly as the $p-d$ band overlap increases.
Due to the $C_{3z}$ protection~\cite{wang2012, yang2014classification}, this band inversion leads to a pair of stable Dirac points rather than a nontrivial band gap on the $\Gamma-A$ line as shown in Fig.~\ref{fig:fig2}(e).
Only when $y$ exceeds 0.32, the $d_{\pm 1/2}$ states could invert with the $p_{\pm 1/2}$ states, and a nontrivial band gap can be opened, leading to a topological metal phase as shown in Fig.~\ref{fig:fig2}(f).

\emph{Effective model} ---
Based on the DFT results, the bases describing the low energy bands along the $\Gamma - A$ path can be simplified as
\begin{equation} \label{eq:eq1}
\begin{split}
    |\phi_{1}\rangle_{+} & = (d_{xz} + i d_{yz})/\sqrt{2}  \\
    |\phi_{2}\rangle_{+} & = (d_{xz} - i d_{yz})/\sqrt{2}  \\
    |\phi_{3}\rangle_{-} & = (p_x  + i p_y)/\sqrt{2}   \\
    |\phi_{4}\rangle_{-} & = (p_x  - i p_y)/\sqrt{2}  , \\
\end{split}
\end{equation}
where the subscripts $\pm$ denote the parity of the basis.
As we discussed above, the band inversion between $|\phi_{1,2}\rangle$ and $|\phi_{3,4}\rangle$ along the $\Gamma - A$ path could lead to the topologically nontrivial band structures in 1$T$-TiTe$_2$.
The effective model at the $\Gamma$ point has the full point group symmetry $D_{3d}$ of the crystal with the generator of rotation $C_{3z}$,  inversion $I$ and mirror $M_x$.
Under the constrain of $D_{3d}$ and time-reversal symmetry, the effective Hamiltonian without SOC is restricted to the following form:
\begin{align}
H_{0}(k) =
\left(
\begin{array}{cccc}
    M_d(k)   &   0    &   it_{1} f(k_z)  & t_{2} k_{+}    \\
    0  &  M_d(k)    &   -t_2 k_{-}  &  it_1 f(k_z) \\
    -it_1 f(k_z) &  -t_2 k_{+} &   M_p(k)  &  0  \\
    t_2 k_{-} & - it_1 f(k_z) &  0 &  M_p(k) \\
\end{array}
\right),
\end{align}
with
\begin{equation}
M_{i} = E_{i} + t_{i}^{\varparallel}(k_{x}^2+k_{y}^2) + t_{i}^{z} g(k_z), \quad i = d, p,
\end{equation}
and
\begin{equation}
\begin{split}
    f(k_z) & = \sin(ck_z),  \\
    g(k_z) & = 1-\cos(ck_z). \\
\end{split}
\end{equation}
where, $E_{d,p}$ are the band energies of $d_{xz \& yz}$ and $p_{x \& y}$ orbitals at the $\Gamma$ point,
$t_{i}^{\varparallel}$ and $t_{i}^z$ are the in-plane and $z$-direction hopping amplitudes of the $i$-th band.
$c$ is the lattice constant along the $z$ direction.
$t_1$ and $t_2$ are the couplings between $p$ and $d$ orbitals.

\begin{table}[!t]
    \caption{Parameters in the effective model.
    }
 \label{table:table1}
 \begin{tabular}{c|c|c|c|c|c}
\toprule
\hline\hline
     $E_d$ (eV) & $E_p$ (eV) & $t_1$ (eV) &  $t_2$ (eV$\cdot${\AA}) &  $t_{d}^{\varparallel}$ (eV$\cdot${\AA}$^2$) & $t_p^{\varparallel}$ (eV$\cdot${\AA}$^2$)  \\
\hline
     0.324  & 0.789 &  0.046  &   3.1  & 7.5  & $-$7.0 \\
\hline
\hline
     $t_d^z$(eV)  &  $t_{p}^z$ (eV)  & $\lambda_d$ (eV) &  $\lambda_p$ (eV) &  $\lambda_1$ (eV) & $\lambda_2$(eV$\cdot${\AA}) \\
\hline
     0.001 & $-$0.211 &  0.094 & 0.154 & 0.113 & 2.0 \\
  \hline
\bottomrule
\end{tabular}
\end{table}

The full Hamiltonian with SOC takes the form $H(k) = I \otimes H_0(k) + H_{\text{SOC}}$
under the spinful bases $(|\uparrow\rangle,|\downarrow\rangle)\otimes(|\phi_{1}\rangle, |\phi_{2}\rangle, |\phi_{3}\rangle, |\phi_{4}\rangle)$.
The symmetry allowed $H_{\text{SOC}}$ is given explicitly by Eqs. S1 - S3 (See details in Section I of Supplemental Material (SM)~\cite{supply}).
Four new parameters induced by the SOC interaction are considered.
$\lambda_{d,p}$ are the on-site SOC strengths for $d$ and $p$ orbitals, respectively.
$\lambda_1$ is the first order SOC induced by the $z$-direction hopping of the $p$ orbitals,
and $\lambda_2$ is induced by the in-plane hopping between the $p$ and $d$ orbitals with opposite spin~\cite{supply}.

\begin{figure}
\includegraphics[width=0.48\textwidth]{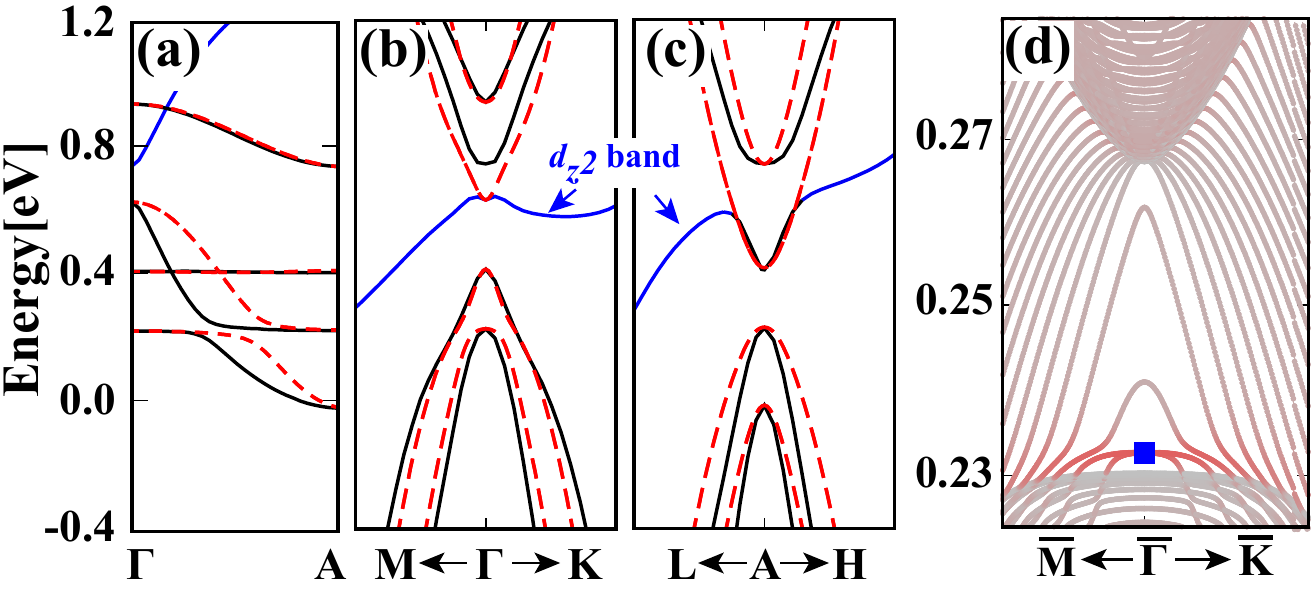}
\caption{
   (a) -- (c) show the fitted band structures from the effective model (dashed red bands) with the DFT calculations (solid black and blue bands) along high-symmetry $k$-path.
   The blue bands are mostly contributed by the $d_{z^2}$ orbital, which are not considered in our effective model.
   (d) The (001) surface states calculated by the effective Hamiltonian, in which the blue square denotes the position of the 2D Dirac cone.
   In all figures, the Fermi level is set at 0 eV.
   }
\label{fig:fig3}
\end{figure}

We use the effective model to fit the DFT calculated band structures of 1$T$-TiTe$_2$, and list the fitted parameters in Table~\ref{table:table1}.
The fitted band structures (red) are plotted in Fig.~\ref{fig:fig3}(a)--(c),
which show that our model and parameters successfully capture the band dispersions and topological characters of the 1$T$-TiTe$_2$ along the $\Gamma - A$ path.
Especially, a topologically nontrivial band gap about 36 meV is opened, and a Dirac point presents at about 0.4 eV above the Fermi level, as shown in Fig.~\ref{fig:fig3}(a).
As shown in Fig.~\ref{fig:fig3}(b) and~\ref{fig:fig3}(c), the in-plane band dispersions below 0.5~eV are reproduced reasonably well.
The mismatch parts are mainly contributed by the $d_{z^2}$ orbital (blue bands in Figs.~\ref{fig:fig3}), which has not been considered in our effective model.
However, we notice that the topological properties of the band structures are not determined by the $d_{z^2}$ band.
It also does not influence the phase diagram of the topological superconductivity,
because which is mainly dominated by the low energy physics at the $\Gamma$ and  $A$ points~\cite{PhysRevLett.107.097001, PhysRevLett.117.047001}.

We perform the surface states calculation based on our effective model, and plot the calculated results of the (001) surface in Fig.~\ref{fig:fig3}(d),
in which the weights contributed by the surface layer are highlighted by the red dots.
One can see clearly that a 2D Dirac cone formed by the topological surface states exists between 0.23 and 0.27~eV.
In Fig.~S1~\cite{supply}, we plot the real space charge density distribution of the surface states as illustrated by the blue square in Fig.~\ref{fig:fig3}(d),
which are mainly distribute on the surface within 10 $\AA$ and shows an exponential decay with the depth.
These results confirm the topological electronic properties of our model in TiTe$_2$ successfully.

\emph{Topological superconductivity} ---
Considering 1$T$-TiTe$_2$ is a superconductor under the pressure of about 5$-$12~GPa,
we would like to study the possible TSC phase realized in it in the following.
Since the crystal structure and the topological band structures along $\Gamma - A$ changes slightly under pressure~\cite{PhysRevB.88.155317},
the parameters in Table~\ref{table:table1} are used to describe the band structures of 1$T$-TiTe$_2$ under pressure.
Similar to Fe(Se,Te)~\cite{PhysRevLett.117.047001}, 1$T$-TiTe$_2$ is a type-II superconductor.
When the superconducting 1$T$-TiTe$_2$ is in an external magnetic field,
the magnetic flux would penetrate into the superconductor and form many vortex lines in the sample ~\cite{Dutta2018}.
Thus, one can view the vortex line as a Majorana chain, and study its BdG spectrum to determine the whole system's topological properties~\cite{Kitaev_2001, PhysRevLett.107.097001}.

Since the magnetic field is usually very weak in the vortex lines, we ignore the vector potential ${\bm A}(r)$
and the zeeman effect in the type-II limit~\cite{CAROLI1964307, PhysRevB.41.822}.
Then the BdG Hamiltonian of the vortex can be written as
\begin{align}\label{eq:eq5}
    H_{\text{BdG}} (r, \theta) =
    \left(
    \begin{array}{cc}
        H'(k) - \mu & \Delta e^{-i\theta}\tanh(r/\xi)\\
        \Delta e^{ i\theta }\tanh(r/\xi) & -H'^\ast(-k) + \mu \\
    \end{array}
    \right).
\end{align}
Here the vortex is assumed along the z-direction.
$H'(k)$ is the 8-band Hamiltonian obtained from $H(k)$ by a unitary transformation~\cite{supply}.
$\mu$ is the chemical potential, $(r,\theta)$ are the polar coordinates in the $\text{xy}$-plane.
$\xi=3.6$ nm is the Ginzburg-Landau coherent length~\cite{PhysRevB.97.060503}.
A uniform $s$-wave superconducting pairing is adopted by assuming
$\Delta = \Delta_0(i \tau_y \otimes I_4)$ with $\Delta_0 = 1.2$~meV~\cite{PhysRevB.97.060503},
where $I_4$ is a $4\times 4$ identity matrix, and $\tau_y$ is the second Pauli matrix.

\begin{figure}[tbp]
\includegraphics[clip,width=0.49\textwidth, angle=0]{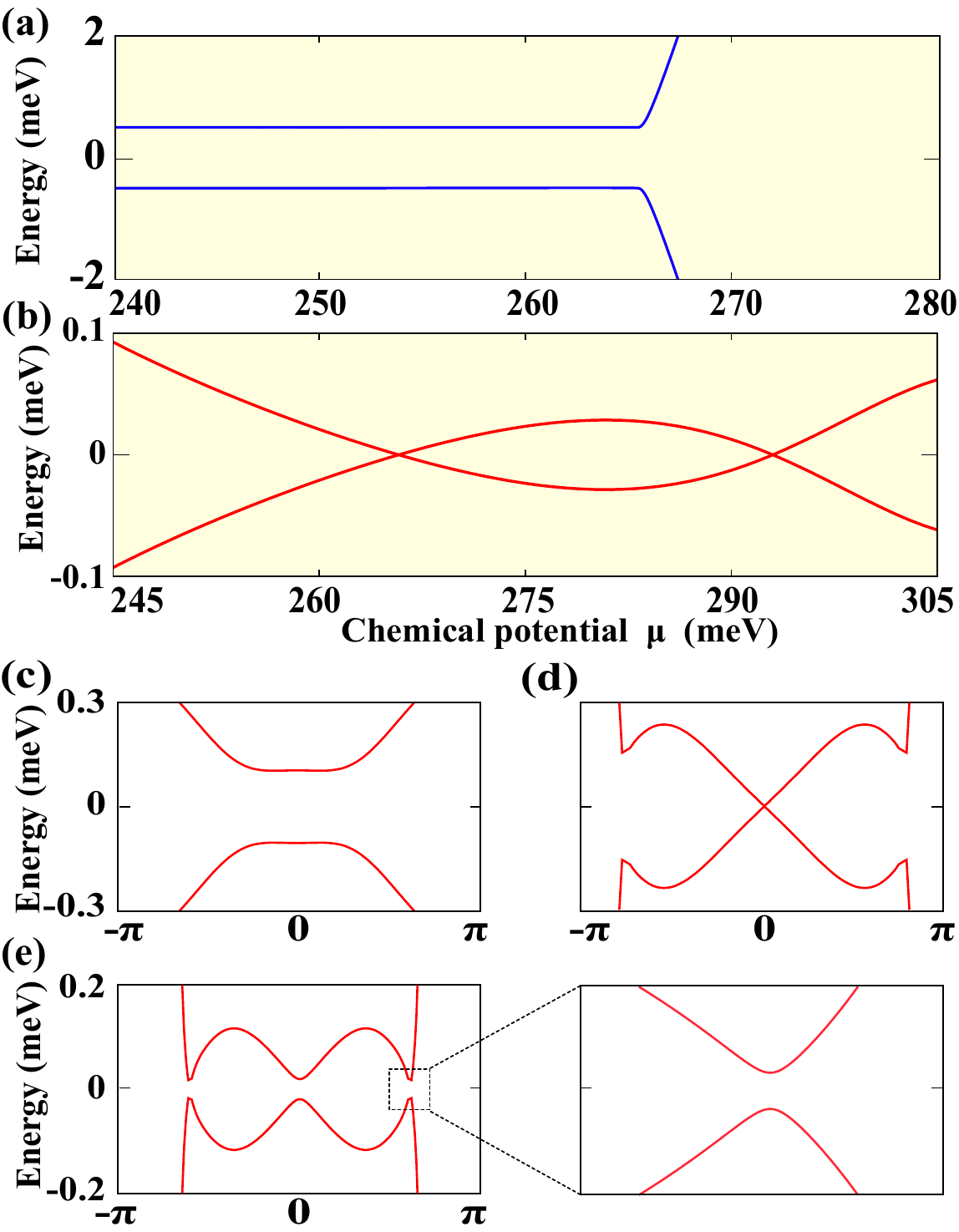}
\caption{(a) and (b) are the energy spectrum of the vortex line at the $A$ and the $\Gamma$ point as a function of chemical potential $\mu$, respectively.
While the spectrum is always gapped at $A$ point, the energy spectrum at $\Gamma$ point shows two gap closing at the critical chemical potentials $266$ and $293$~meV.
(c) -- (e) are the low energy dispersions of the BdG Hamiltonian of the vortex line for the NSC phase (c with $\mu =$~245 meV),
the critical point of phase transition (d with $\mu =$~266 meV ) and the TSC phase (e with $\mu =$~280~meV), respectively.
}
\label{fig:fig4}
\end{figure}

For the Majorana chain described by Eq.~\ref{eq:eq5}, only $k_z$ is the good quantum number,
and its topological phase transition only be characterized by gap closing of the BdG spectrum at either $k_z = 0$ ($\Gamma$) or $k_z = \pi$ ($A$)~\cite{PhysRevLett.107.097001}.
Therefore, by discretizing the polar coordinate r on a cylinder with the radius 3.6~$\mu$m, we numerically solve the vortex BdG Hamiltonian,
and plot the spectrum at the $A$ and the $\Gamma$ point as a function of the chemical potential $\mu$ in Fig.~\ref{fig:fig4}(a) and~\ref{fig:fig4}(b), respectively.
The calculated results obviously manifest that the spectrum at the $A$ point is always gapped,
while the energy spectrum at the $\Gamma$ point closes its gap at the critical chemical potentials $\mu_{c1} = 266$~meV and $\mu_{c2} = 293$~meV.
Considering the bulk pairing dominates at the chemical potential far away from the Dirac cone,
our results strongly suggest that the vortex line falls into the normal superconductivity (NSC) phase both for $\mu < 266$~meV and $\mu > 293$~meV,
and the TSC phase is realized in the chemical potential interval $\mu \in [\mu_{c1}, \mu_{c2}]$.
We note that the energy interval $\mu \in [\mu_{c1}, \mu_{c2}]$ is not exactly corresponds to the energy window of the surface Dirac cone shown in Fig. ~\ref{fig:fig3}(d).
Such shift have been reported and explained in previous literatures~\cite{Chiu2012}.

In Figs.~\ref{fig:fig4}(c)-(e) we plot the BdG spectrum of the vortex line at different chemical potentials in the NSC phase [$\mu =$~245 meV for Fig.~\ref{fig:fig4}(c)],
at the phase transition point [$\mu = 266$ meV for Fig.~\ref{fig:fig4}(d)] and in the TSC phase [$\mu =$~280 meV for Fig.~\ref{fig:fig4}(e)], respectively.
They clearly show a gap closing and reopening process at $\Gamma$ point as $\mu$ varied.
In particular, the spectrum beside the transition point is fully gapped at all $k_z$, which is the crucial requirement to protect stable MZMs at the vortex ends.
As we all know, Zak phase is a well-defined topological number to characterize the topological property of the 1D superconductor as the gap opened~\cite{PhysRevLett.62.2747, PhysRevB.73.245115}.
We notice that, different from the Fe(Se,Te), the gap closing point of chemical potential in 1$T$-TiTe$_2$ is not affected by the bulk superconducting gap $\Delta_0$, as shown in Fig.~S2(a).
More importantly, the gap size at the fixed $\mu$ is increased monotonously with $\Delta_0$ as shown in Fig.~S2(b).
Taking advantages of these properties, we can calculate the Zak phase at large $\Delta_0$ to determine the real TSC region in 1$T$-TiTe$_2$.
We have calculated the evolution of the Zak phase as a function of $\mu$ with $\Delta_0 = 50$ meV,
which manifests that the Zak phase is 0 at beginning ($\mu < 267$~meV), and raises to $\pi$ at $\mu \in $ [267~meV, 291~meV], and finally turns back to 0 as $\mu > 291$~meV as shown in Fig.~S2(c).
These results verify that the vortex chain in 1$T$-TiTe$_2$ falls into the TSC phase as $\mu \in [\mu_{c1}, \mu_{c2}]$~\cite{PhysRevLett.107.097001, PhysRevLett.117.047001}.
Our results also suggest that, similar to the TSC phase in Cu$_x$Bi$_2$Se$_3$ ~\cite{PhysRevLett.100.096407, PhysRevLett.107.097001},
one can improve the stability of the TSC phase and enhance the alive temperature of MZMs in 1$T$-TiTe$_2$ by increasing its bulk superconducting gap,
which is another advantage that the TSC phase in iron-based superconductors do not have.

\emph{Discussion} ---
Finally, we would like to discuss another highly possible route to realize TSC in titanium dichalcogenide.
As we all know, some transition metal intercalation can also suppress the charge density wave transition, and induce the superconductivity in 1$T$-TiSe$_2$.
For example, $T_c=4$~K superconductivity transition has been observed in the Cu intercalated 1$T$-TiSe$_2$~\cite{NaturePhysics.2.544}.
According to our calculations, the Te doped 1$T$-TiSe$_2$ have very similar band structures with TiTe$_2$,
including the $p-d$ band inversion, a 2D Dirac cone and a topological gap as shown in Fig.~S3.
Moreover, such Te doping can enhance the density of states at Fermi level, which is beneficial for superconductivity~\cite{shang2019}.
Therefore, one can expect that TSC and MZMs can be realized in the Te doped Cu$_x$TiSe$_2$ system.
Taking the half doped 1$T$-TiSe$_2$ as a concrete example, we have calculated its vortex BdG model with $\Delta_0 = 0.8$~meV, and plot the spectrum in Fig.~S4,
which exhibits similar spectrum gap closing behavior as in 1$T$-TiTe$_2$, indicating that Te doped 1$T$-TiSe$_2$ is also a promising TSC candidate.

\emph{Acknowledgements.} ---
This work is supported by the National Key R \& D Program of China (No.~2018YFA0307000) and the National Natural
Science Foundation of China (No.~11874022).

\onecolumngrid
\clearpage
\begin{center}
\textbf{\large Supplemental Materials: Prediction of topological superconductivity in 1$T$-TiTe$_2$ under pressure}
\end{center}

\setcounter{equation}{0}
\setcounter{figure}{0}
\setcounter{table}{0}
\setcounter{page}{1}
\makeatletter

\renewcommand{\thefigure}{S\arabic{figure}}
\renewcommand{\thetable}{S\arabic{table}}
\def\theequation{S\arabic{equation}}

\section{Section I: Effective model and edge states}

\par
Based on the DFT results, the bases describing the low energy bands along the $\Gamma - A$ path can be simplified as Eq.~1 in the main text.
So the full Hamiltonian with SOC takes the form $H(k) = I \otimes H_0(k) + H_{\text{SOC}}$
under the spinful bases $(|\uparrow\rangle,|\downarrow\rangle)\otimes(|\phi_{1}\rangle, |\phi_{2}\rangle, |\phi_{3}\rangle, |\phi_{4}\rangle)$.
The symmetry allowed $H_{\text{SOC}}$ is given explicitly by Eqs.~S1--S3:
\begin{align}
h_{\text{SOC}}(k) =
\left(
\begin{array}{cc}
\Lambda_{\uparrow\uparrow} & \Lambda_{\uparrow\downarrow}\\
\Lambda^\dag_{\uparrow\downarrow} & \Lambda_{\downarrow\downarrow}\\
\end{array}
\right),
\end{align}
with
\begin{align}
\Lambda_{\uparrow\uparrow} = -\Lambda_{\downarrow\downarrow} =
\left(
\begin{array}{cccc}
-\lambda_d & 0& 0& 0 \\
0 &  \lambda_d & 0 & 0 \\
0 & 0 & - \lambda_p -\lambda_1 g(k_z)  & 0 \\
0 & 0 & 0  &  \lambda_p + \lambda_1 g(k_z) \\
\end{array}
\right),
\end{align}
and
\begin{align}
\Lambda_{\uparrow\downarrow} =
\left(
\begin{array}{cccc}
0 & 0& 0& 0 \\
0 &  0 & i\lambda_2 k_+ & 0 \\
0 & 0 & 0  & 0 \\
i\lambda_2 k_+ & 0 & 0 & 0 \\
\end{array}
\right).
\end{align}
$\lambda_{d,p}$ are the on-site SOC strengths for $d$ and $p$ orbitals, respectively.
$\lambda_1$ is the first order SOC induced by the $z$-direction hopping of the $p$ orbitals,
and $\lambda_2$ is induced by the in-plane hopping between the $p$ and $d$ orbitals with opposite spin.
$g(k_z)$ is defined in the Eq.~(4) of main text.

\par
The $H'(k)$ used in Eq.~(5) of the main text is the 8-band Hamiltonian under
new basis ordering ( $|\phi_{1},\uparrow\rangle, |\phi_{2},\uparrow\rangle, |\phi_{3},\uparrow\rangle, |\phi_{4},\uparrow\rangle,
|\phi_{2},\downarrow\rangle, |\phi_{1},\downarrow\rangle, |\phi_{4},\downarrow\rangle, |\phi_{3},\downarrow\rangle $ ),
which is obtained from $H(k)$ by a unitary transformation as following:

\begin{align}
H'(k)= UH(k)U^\dag,
\end{align}
where $U$ is a unitary matrx:
\begin{align}
U= \left(
\begin{array}{cccccccc}
1 & 0 & 0 & 0 & 0 & 0 & 0 & 0\\
0 & 1 & 0 & 0 & 0 & 0 & 0 & 0\\
0 & 0 & 1 & 0 & 0 & 0 & 0 & 0\\
0 & 0 & 0 & 1 & 0 & 0 & 0 & 0\\
0 & 0 & 0 & 0 & 0 & 1 & 0 & 0\\
0 & 0 & 0 & 0 & 1 & 0 & 0 & 0\\
0 & 0 & 0 & 0 & 0 & 0 & 0 & 1\\
0 & 0 & 0 & 0 & 0 & 0 & 1 & 0\\
\end{array}
\right).
\end{align}
\par
In order to check the topological electronic properties of our model in $1T$-TiTe$_2$,
we plot the real space charge density distribution of the surface state as illustrated by the blue square in Fig.~3(d) of the main text.
As shown in Fig.~S1, the charge density of the 2D Dirac cone are mostly accumulated at the system surface.
As a result, we have sufficient evidence to point out that the 2D Dirac cone is formed by topological surface states.

\begin{figure}[!t]
\includegraphics[width=0.5\columnwidth]{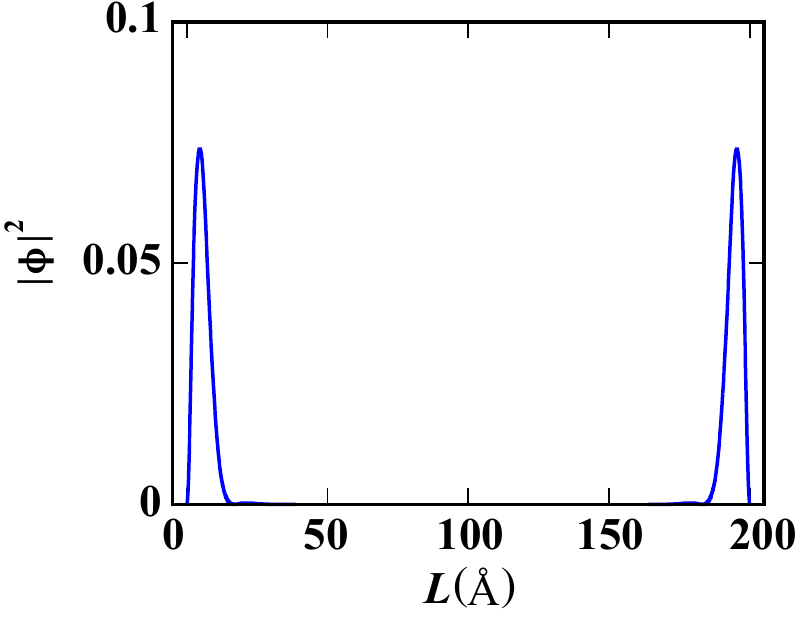}
\caption{The charge distribution in real space for the 2D Dirac cone that illustrated by the blue square in Fig.3(d) of the main text.}
\label{fig:figs1}
\end{figure}

\section{Section II : TSC phase region and Zak phase}
\par
In Fig.~S2(a), we plot the energy spectrum at the $\Gamma$ point as a function of the chemical potential $\mu$ for different bulk
superconducting gap $\Delta_0$. Obviously, we find that the increase of $\Delta_0$ does not change the region of the TSC phase. In Fig.~S2(b),
we show the low energy dispersions of the BdG Hamiltonian with different $\Delta_0$ at $\mu = 280$~meV,
which demonstrates that the gap size on the vortex line is increased monotonously as $\Delta_0$ increases.
In Fig.~S2(c), we calculate the evolution of the Zak phase $\phi$ as a function of $\mu$ with $\Delta_0 = $ 50 meV
for researching the topological property of the vortex line,
which manifests that $\phi=0$ at $\mu < 267$~meV,
$\phi=\pi$ at $\mu \in $ [267 meV, 291 meV],
and $\phi=0$ at $\mu > 291$ meV, respectively.
This result confirm the topologically nontrivial region at $\mu \in $ [267 meV, 291 meV].

\begin{figure}[!b]
\includegraphics[width=\columnwidth]{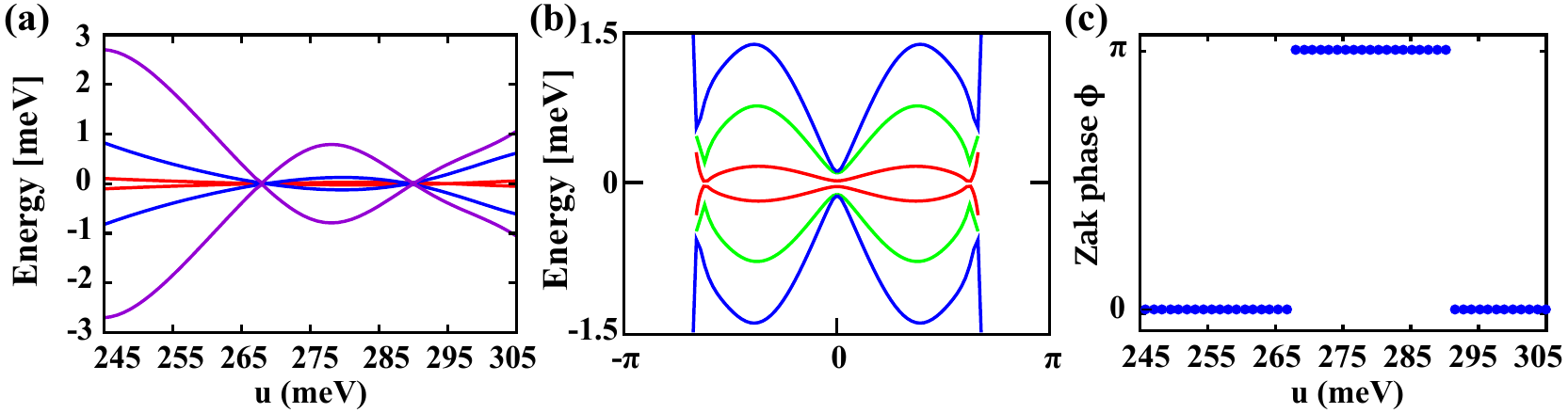}
\caption{
(a) The low-energy spectrum at the $\Gamma$ point with $\Delta_0$ =
1 (red), 10 (blue),  50 (violet) meV, respectively.
In particular, we note that the energy spectrum calculated by $\Delta_0$ = 50 meV is multiplied 1/5.
(b) The energy dispersions of the BdG Hamiltonian at  $\mu = 280$~meV with $\Delta_0$ =
1 (red), 5 (green),  10 (blue) meV, respectively.
(c) The evolution of the Zak phase as a function of $\mu$ with $\Delta_0 = $ 50 meV,
which manifests that the Zak phase $\phi=\pi$ at $\mu \in $ [267 meV, 291 meV].
}
\label{fig:figs2}
\end{figure}

\section{Section III : Band structure and TSC in Ti(Se$_{0.5}$Te$_{0.5}$)$_2$ $ $ }

\par
In this section, we calculate the band structures of Ti(Se$_{0.5}$Te$_{0.5}$)$_2$ as shown in Fig.~S3(a),
which are very similar to the results of $1T$-TiTe$_2$. Then we use the effective model to fit the DFT calculated band structures of Ti(Se$_{0.5}$Te$_{0.5}$)$_2$,
and list the fitted parameters in Table~S1.
The fitted band structures (red) are plotted in Fig.~S3(b)--(d).
The Fig.~S3(b) shows that our model and parameters successfully capture the band dispersions and energy position of the topologically nontrivial gap in Ti(Se$_{0.5}$Te$_{0.5}$)$_2$ along the $\Gamma-A$ path.
Fig.~S3(c)-(d) show the in-plane band dispersions below 0.5~eV are reproduced reasonably well.

\begin{table}[!b]
\renewcommand*{\arraystretch}{1.5}
\caption{Parameters used for Ti(Se$_{0.5}$Te$_{0.5}$)$_2$.
}
 \label{table:table1}
 \begin{tabular}{c|c|c|c|c|c}
\hline\hline
     $E_d$ (eV) & $E_p$ (eV) & $t_1$ (eV) & $t_2$ (eV$\cdot${\AA})  &  $t_{d}^{\varparallel}$ (eV$\cdot${\AA}$^2$) & $t_p^{\varparallel}$ (eV$\cdot${\AA}$^2$)  \\
\hline
     0.418 & 0.917 &  0.038  &  3.0  & 12.0  & $-$10.0 \\
\hline
\hline
     $t_d^z$(eV)  &  $t_{p}^z$ (eV)  & $\lambda_d$ (eV) &  $\lambda_p$ (eV) &  $\lambda_1$ (eV) & $\lambda_2$(eV$\cdot${\AA}) \\
\hline
     0.004 & $-$0.236 &  0.082 & 0.135 & 0.094 & 2.0 \\
  \hline
\end{tabular}
\end{table}

\begin{figure}[!b]
\includegraphics[width=0.6\columnwidth]{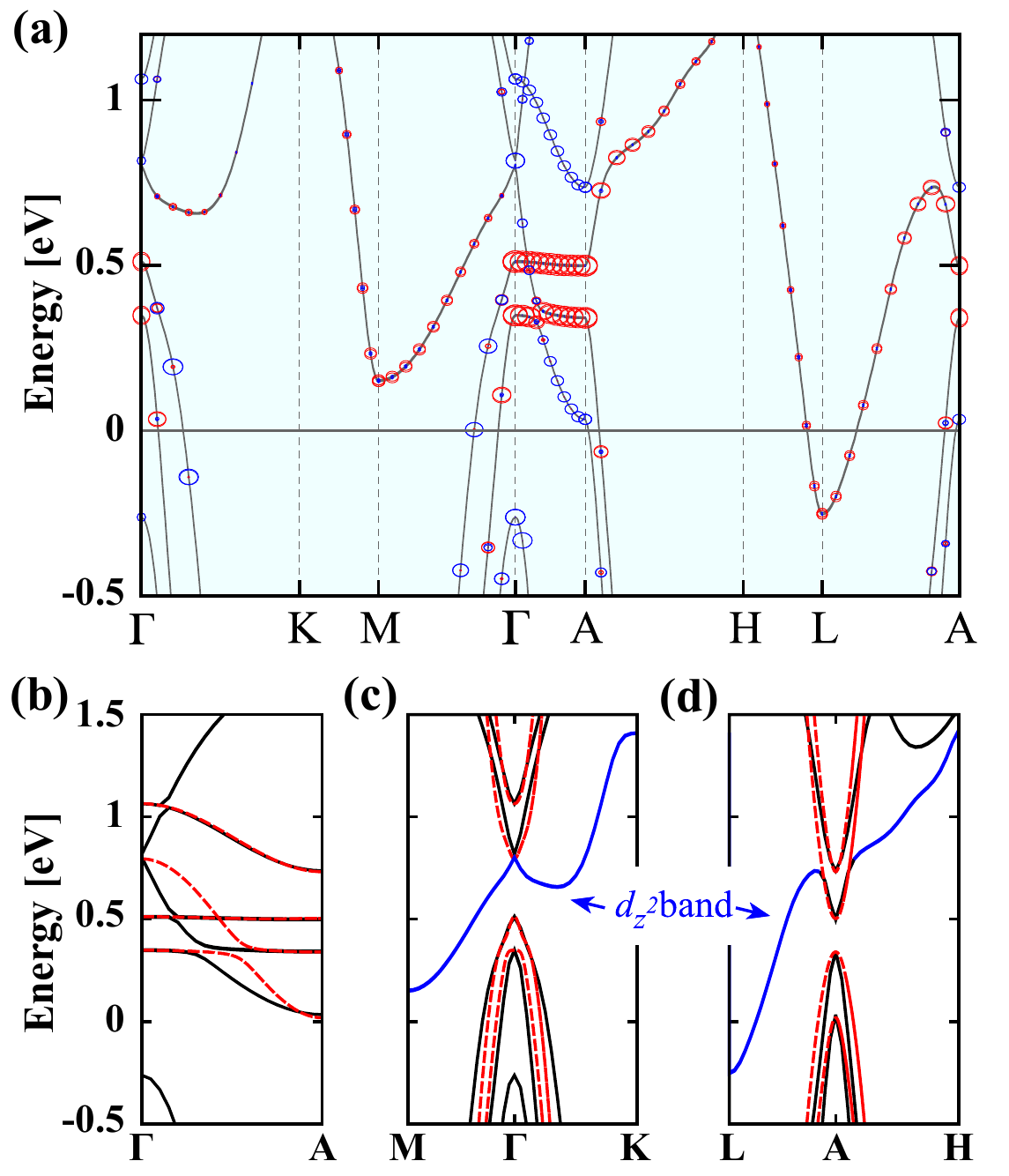}
\caption{
     Band structures of Ti(Se$_{0.5}$Te$_{0.5}$)$_2$ $ $. (a) the HSE06 calculations.
    The size of red and blue circles represent the weight projections for the $d$ orbitals of Ti atoms and the $p$ orbitals of chalcogen atoms, respectively.
    (b) - (d) show the fitted band structures from the effective model (dashed red bands) with the DFT calculations (solid black and blue bands) along high-symmetry $k$-path.
   The blue bands are mostly contributed by the $d_{z^2}$ orbital, which are not considered in our effective model.
}
\label{fig:figs3}
\end{figure}

\par
Finally, we calculate its vortex BdG model with $\Delta_0 = $ 0.8 meV and plot the spectrum in Fig.~S4.
The calculated results obviously manifest that the spectrum at the $A$ point is always gapped,
while the energy spectrum at the $\Gamma$ point closes its gap both at the critical chemical potentials $\mu_{c1} = 348.785$~meV and $\mu_{c2} = 397.6$~meV.
These results exhibit similar spectrum gap closing behavior as shown in 1$T$-TiTe$_2$, indicating that Te doped 1$T$-TiSe$_2$ is also a promising TSC candidate.

\begin{figure}[!t]
\includegraphics[width=0.9\columnwidth]{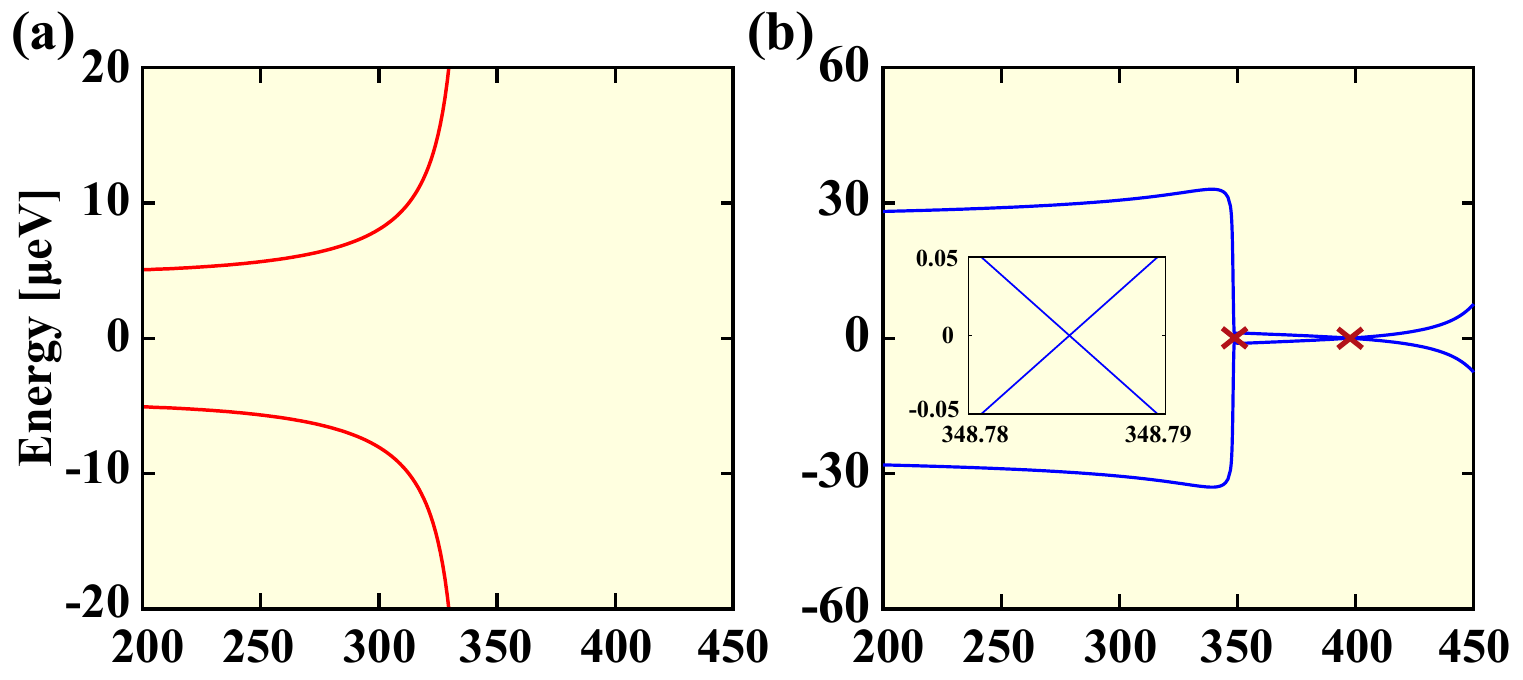}
\caption{
(a) and (b) are the BdG spectrum of Ti(Se$_{0.5}$Te$_{0.5}$)$_2$ with respect to chemical potential $\mu$ at the $A$ and the $\Gamma$ point, respectively.
The spectrum is always gapped at the $A$ point. At the $\Gamma$ point, two critical chemical potentials at about $348.785$ and $397.6$~meV
are marked by red crossing. The inset panel shows the detailed spectrum around $348.785$~meV.
}
\label{fig:figs4}
\end{figure}

\end{document}